\documentclass{appolb}

\usepackage{amsmath, amsfonts, bm, caption, geometry, graphicx, hyperref, subcaption, xspace}
\usepackage{slashed}
\usepackage{subcaption}

\usepackage{pdflscape}

\usepackage{color,colordvi}

\DeclareMathOperator{\Order}{\mathcal{O}}

\newcommand{\D}{\mathrm{d}}

\newcommand{\ep}{\epsilon}
\newcommand{\eps}{\epsilon}
\newcommand{\nn}{\nonumber}

\newcommand{\process}{e^+ + e^- \to \gamma^* \to g + \langle\text{$n$ partons}\rangle}
\newcommand{\sd}[2]{\,#1\!\cdot\!#2}

\newcommand{\FIRE}{\texttt{FIRE}\xspace}
\newcommand{\FORM}{\texttt{FORM}\xspace}
\newcommand{\HPL}{\texttt{HPL}\xspace}
\newcommand{\IBP}{IBP \cite{CT81,Tka81}\,}
\newcommand{\LiteRed}{\texttt{LiteRed}\xspace}
\newcommand{\Mathematica}{\texttt{Mathematica}\xspace}
\newcommand{\QGRAF}{\texttt{QGRAF} \cite{Nog91}\,}
\newcommand{\Reduze}{\texttt{Reduze}\xspace}
\newcommand{\harmpol}{\texttt{harmpol}\xspace}

\def\Fa{{\cal F}_A}
\def\Fl{{\cal F}_L}
\def\Ft{{\cal F}_T}

\def\Gijz{P_{ij}^{(0)}}

\def\Giqz{P_{iq}^{(0)}}

\def\Gjqz{P_{jq}^{(0)}}

\def\Gpiz{P_{pi}^{(0)}}
\def\Gpqz{P_{pq}^{(0)}}

\def\Giqo{P_{iq}^{(1)}}
\def\Gpio{P_{pi}^{(1)}}
\def\Gpqo{P_{pq}^{(1)}}

\def\Gpqt{P_{pq}^{(2)}}

\def\Cio{c_{i}^{(1)}}
\def\Cjo{c_{j}^{(1)}}

\def\Cit{c_{i}^{(2)}}
\def\Cpt{c_{p}^{(2)}}

\def\Cpd{c_{p}^{(3)}}

\def\MSbar{\overline{\text{MS}}}

\def\Aio{a_{i}^{(1)}}
\def\Ajo{a_{j}^{(1)}}

\def\Ait{a_{i}^{(2)}}
\def\Apt{a_{g}^{(2)}}

\def\Bio{b_{i}^{(1)}}
\def\Bjo{b_{j}^{(1)}}

\def\btio{b_{i}^{(1)}}

\begin{document}
  \eqsec  % uncomment this line to get equAions numbered by (sec.num)
  
  \title{
  \begin{flushright}
    DESY 15-064 \\
    IFJPAN-IV-2015-7 \\
  \end{flushright}
  Towards three-loop QCD corrections \\ to the time-like splitting functions
  \thanks{Presented by O.Gituliar at XXI Cracow Epiphany Conference, Cracow (Poland), 8--10 Jan 2015.}
  }
  
\author{O.~Gituliar$^{\,a}$, S.~Moch$^{\,b}$
\address{$^a$ Institute of Nuclear Physics, Polish Academy of Sciences \\
              Radzikowskiego 152, 31-342 Cracow, Poland}
\address{$^b$ II. Institute for Theoretical Physics, Hamburg University \\
              D-22761 Hamburg, Germany}
}

\maketitle

\begin{abstract}
We report on the status of a direct computation of the time-like splitting functions at next-to-next-to-leading order in QCD.
Time-like splitting functions govern the collinear kinematics of inclusive hadron production and the evolution 
of the parton fragmentation distributions.
Current knowledge about them at three loops has been inferred by means of crossing symmetry from their related space-like counterparts, 
which has left certain parts of the off-diagonal quark-gluon splitting function undetermined. 
This motivates an independent calculation from first principles. 
We review the tools and methods which are applied to attack the problem.
\end{abstract}


\section{Introduction}

Splitting functions are universal quantities in QCD. 
They govern the collinear evolution in hard scattering processes with hadrons in the initial or final state.
In reactions with initial state protons the parton luminosity is parametrized by parton distribution functions
whose scale dependence is subject to evolution equations 
with space-like kinematics (a space-like hard scale $-q^2 \ge 0$) and splitting functions $P^{S}_{ab}(x)$, 
where $x$ denotes the parton's momentum fraction of the proton momentum. 
Similarly, for processes with identified hadrons in the final state the parton to hadron transition 
is described by the parton fragmentation distributions $D_{f}^{h}(x,q^2)$, 
where $x$ represents the fractional momentum of the final-state parton $f$ 
transferred to the outgoing hadron $h$ and $q^2 \ge 0$ is a time-like hard scale.
The scale dependence of the fragmentation distributions 
is controlled by the so-called time-like splitting functions $P^{T}_{ba}(x)$,
and is given by
\begin{eqnarray}
\label{eq:Devol}
  {d \over d \ln q^2} \; D_{a}^{h} (x,q^2) & = &
  \int_x^1 {dz \over z} \,P^{T}_{ba} \left( z, \alpha_s(q^2) \right) D_{b}^{h} \Big(\, {x \over z},\, q^2 \Big) \; ,
\end{eqnarray}
where the summation runs over the number $n_f$ of effectively massless quark flavors and the gluon, 
$b = q_i,\bar{q}_i, g$ for $i = 1, \ldots, n_f$. 

The time-like splitting functions $P^{T}_{ba}$ can be computed in perturbation
theory in powers of the strong coupling $\alpha_s$,
\begin{eqnarray}
\label{eq:PTexp}
  P^{T}_{ba} \left( x,\alpha_s (q^2) \right) & = &
  a_s \, P_{ba}^{(0)T}(x) 
  + a_s^{2}\, P_{ba}^{(1)T}(x)
  + a_s^{3}\, P_{ba}^{(2)T}(x) + \ldots \, ,
\end{eqnarray}
where we normalize the expansion parameter as $a_s = \alpha_s/ (4\pi)$.
At one and two loops, the leading (LO) and next-to-leading order (NLO) splitting functions 
$P_{ba}^{(0)T}$ and $P_{ba}^{(1)T}$ have been obtained with different methods in the past, e.g., 
from the collinear singularities in inclusive hadron production 
in electron-positron annihilation~\cite{Rijken:1996ns,Mitov:2006wy}.

In contrast, the next-to-next-to-leading order (NNLO) terms $P^{(2)T}_{ba}$ at three loops 
have been determined by crossing relations from the respective functions $P^{S}_{ab}$ for 
space-like kinematics based on analytic continuation in the scaling variable $x$, 
i.e., mapping $x \to 1/x$~\cite{Mitov:2006ic,Moch:2007tx,Almasy:2011eq}.
In fact, the LO space-like and time-like splitting functions are identical (up to transposition), 
a fact known as the Gribov-Lipatov relation~\cite{Gribov:1972ri,Gribov:1972rt}.
At higher orders such relations between the space-like splitting
functions, or their analytic continuations, and their time-like counterparts
do not hold in the usual $\MSbar$ scheme~\cite{Blumlein:2000wh}.

Analytic continuations of the corresponding physical evolution kernels
in the respective kinematics can, however, be used for a constructive approach to the time-like splitting functions and yield relations, 
which fix the components of the matrix $P^{(2)T}_{ba}$ at three loops~\cite{Almasy:2011eq}.
For the diagonal terms $P^{(2)T}_{aa}$ these results agree with assumptions about the universality of evolution kernels \cite{Dokshitzer:2005bf}, 
while for the off-diagonal terms an uncertainty in the time-like quark-gluon splitting function $P^{T}_{qg}$ remains. 
This motivates a direct calculation of the time-like splitting functions in perturbative QCD from a physical process. 
We choose electron-positron annihilation and consider inclusive hadron production through photon exchange 
as well as top-quark mediated decay of the Higgs boson into hadrons in the effective theory~\cite{Moch:2007tx}.

The report is organized as follows.
In Section~\ref{sec:2} we show how to reconstruct time-like splitting functions 
from the bare fragmentation functions in the framework of mass factorization, picking the process $\process$ as an example.
Section~\ref{sec:3} is dedicated to some technical aspects of the calculation.
Here we briefly discuss, in particular, how to perform final-state integration 
with the help of integration-by-parts (IBP) method and how to find master integrals from differential equations.
In addition, to illustrate our approach we provide some examples relevant 
to the calculation of NLO corrections to time-like splitting functions.
We summarize in Section~\ref{sec:4}.

\section{The Set-Up} \label{sec:2}

Let us consider the relevant parton processes in electron-positron annihilation
\begin{eqnarray}
  \label{eq:process}
  e^+ + e^- \to \gamma^*(q) \to p(k_0) + \langle\text{$n$ partons}\rangle\,  ,\\
  \label{eq:process-phi}
  e^+ + e^- \to \phi^*(q) \to p(k_0) + \langle\text{$n$ partons}\rangle\, ,
\end{eqnarray}
with photon ($\gamma$) exchange and Higgs ($\phi$) boson exchange 
in the effective theory and the partons $p = q,{\bar q},g$ with momentum $k_0$.
For the photon-exchange process~(\ref{eq:process}), following the notation in \cite{Nason:1993xx}, the unpolarized differential cross-section in $m=4-2\eps$ dimensions is given by
\begin{equation}
  \frac{1}{\sigma_\text{tot}}
  \frac{\D^2 \sigma}{\D x \, \D \cos\theta}
  =
  \frac{3}{8}(1+\cos^2\theta)\, \Ft(x,\eps)
  +
  \frac{3}{4}\sin^2\theta\, \Fl(x,\eps)
  +
  \frac{3}{4}\cos\theta\,  \Fa(x,\eps),
\end{equation}
where $\theta$ denotes an angle between the beam and parton momentum $k_0$.
The scaling variable $x$ is defined as
\begin{align}
  x=\frac{2\sd{q}{k_0}}{q^2},
  &&
  q^2 = s > 0,
  &&
  0<x\le1\, .
\end{align}
In the center-of-mass frame of the $e^+e^-$ pair $x$ 
can be interpreted as a fraction of the beam energy carried by the parton with momentum $k_0$.
The {\em transverse}, {\em longitudinal}, and {\em asymmetric} fragmentation functions are defined as $\Ft$, $\Fl$, and $\Fa$ respectively.
They are the analog quantities of the deep-inelastic structure functions for space-like $q$ ($-q^2 \ge 0$).

\subsection{Mass factorization}

In the context of the {\em mass factorization} formalism, the transverse fragmentation function $\Ft(x,\eps)$ in the $\MSbar$ scheme can be written as 
\begin{equation}
  {\cal F}_{T,p}(x,\eps) = \sum_{n=0}^{\infty}\, a_s^n\, {\cal F}_{T,p}^{(n)}(x,\eps)\, , 
\qquad p=q,g\, ,
\end{equation}
and explicitly, in terms of coefficients of the QCD $\beta$-function and splitting functions up to NNLO 
for $e^+ + e^- \to \gamma^* \to p(k_0) + \langle\text{$n$ partons}\rangle$ 
\begin{eqnarray}
  {\cal F}_{T,p}^{(1)} \label{eq:ft1}
  & = &
      - \frac{1}{\eps} \Gpqz
    + c_{p}^{(1)}
    + \eps \, a_{p}^{(1)}
    + \eps^2 \, b_{p}^{(1)}
    + \Order(\eps^3),
  \\
  {\cal F}_{T,p}^{(2)} \label{eq:ft2}
  & = &
    \frac{1}{\ep^2}
    \bigg\{
        \frac{1}{2} \Gpiz \Giqz
      + \frac{1}{2} \beta_0 \Gpqz
    \bigg\}
    -
    \frac{1}{\ep}
    \bigg\{
      \frac{1}{2} \Gpqo + \Gpiz \Cio
    \bigg\}
    +
    \bigg\{
      \Cpt - \Gpiz \Aio
    \bigg\}
  \nn \\ & + &
    \ep \,
    \bigg\{
      \Apt - \Gpiz \btio
    \bigg\}
    +
    \Order(\eps^2),
\\
  {\cal F}_{T,p}^{(3)} \label{eq:ft3}
  & = &
  - \frac{1}{\ep^3}
  \bigg\{
      \frac{1}{6} \Gpiz \Gijz \Gjqz
    + \frac{1}{2} \beta_0 \Gpiz \Giqz
    + \frac{1}{3} \beta_0^2 \Gpqz
  \bigg\} 
  \nn \\ & + &
  \frac{1}{\ep^2}
  \bigg\{
      \frac{1}{6} \Gpiz \Giqo
    + \frac{1}{3} \Gpio \Giqz
    + \frac{1}{2} \Gpiz \Gijz \Cjo
    + \frac{1}{3} \beta_1 \Gpqz
    + \beta_0 \left( \frac{1}{3} \Gpqo + \frac{1}{2} \Gpiz \Cio \right)
  \bigg\}
  \nn \\ & - &
  \frac{1}{\ep}
  \bigg\{
      \frac{1}{3} \Gpqt
    + \frac{1}{2} \Gpio \Cio
    - \frac{1}{2} \Gpiz \Gijz \Ajo
    + \Gpiz \Cit
    - \frac{1}{2} \beta_0 \Gpiz \Aio
  \bigg\}
  \nn \\ & + &
  \bigg\{
      \Cpd
    - \Gpiz \Ait
    + \frac{1}{2} \Gpiz \Gijz \Bjo
    - \frac{1}{2} \Gpio \Aio
    + \frac{1}{2} \beta_0 \Gpiz \Bio
  \bigg\}
  +
  \Order(\eps),
\end{eqnarray}
where summation over the repeated indices $i,j=q,g$, Mellin convolution in $x$-space, and the normalization ${\cal F}_{T,p}^{(0)} = \delta(1-x)$ is understood.

The term in eq.~\eqref{eq:ft3} proportional to $1/\eps$ for $p=g$ 
contains the off-diagonal time-like splitting function $P^{(2)T}_{gq}$ (plus some terms of lower orders which are known).
The fragmentation functions ${\cal F}_{\phi,p}^{(i)}$ for the Higgs boson decay 
into hadrons in the effective theory~\cite{Moch:2007tx} with the normalization ${\cal F}_{\phi,g}^{(0)} = \delta(1-x)$, 
i.e. $e^+ + e^- \to \phi^*(q) \to p(k_0) + \langle\text{$n$ partons}\rangle$, 
allow the extraction of $P^{(2)T}_{qg}$ (in which we are interested) from ${\cal F}_{\phi,q}^{(i)}$.
The corresponding expressions for the mass factorization 
can be obtained by replacing $q \to g$ in eqs.~(\ref{eq:ft1}--\ref{eq:ft3}).

\subsection{Perturbative Expansion in QCD}

In perturbative QCD, the transverse fragmentation function $\Ft(x,\eps)$ can be calculated 
with standard means as (see \cite{Rijken:1996ns})
\begin{equation} \label{eq:Ft}
  {\cal F}_T(x,\eps)
   =
   \frac{2}{2-m} \left(
     \frac{\sd{q}{k_0}}{q^2} W_\mu^{\mu} + \frac{k_0^\mu k_0^\nu}{\sd{q}{k_0}}  W_{\mu\nu}
   \right)
,
\end{equation}
where the hadronic tensor $W_{\mu\nu}(x,\eps)$ 
(for the case of photon exchange) reads
\begin{equation} \label{eq:W}
  W_{\mu\nu}(x,\eps) = \frac{x^{m-3}}{4\pi} \int \D \mathrm{PS}(n) \; M_\mu (n) \: M_\nu (n)
  .
\end{equation}
Here $\D \mathrm{PS}(n)$ is $n$-particle real phase-space, defined in eq.~\eqref{eq:psn}, 
and $M_\mu(n)$ describes amplitudes for the process of eq.~\eqref{eq:process}
and the index $\mu$ corresponds to the polarization of the virtual photon,
which is summed over.

To generate Feynman diagrams for the processes \eqref{eq:process} and (\ref{eq:process-phi})
we use \QGRAF 
and process its output further with the help of \FORM \cite{form}, 
i.e., contract indices, calculate Dirac traces and $SU(N)$ color factors, make partial fractioning, etc.
The resulting scalar expression is suitable for the further final-state integration as described in Section~\ref{sec:3}.

In the remaining part of this section let us discuss which contributions to the NNLO splitting functions are known and which actually need to be calculated.

\begin{figure}[h]
  \centering
  \begin{subfigure}[b]{0.3\textwidth}
    \includegraphics[width=\textwidth]{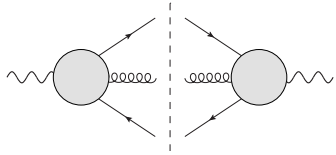}
  \end{subfigure}%
  \caption{Contributions 
    to the time-like splitting function at LO: 
    $P_{gq}^{(0)T}$ 
    (tagged gluon, photon exchange process \eqref{eq:process}); 
    respectively 
    $P_{qg}^{(0)T}$  (tagged quark, Higgs boson exchange process \eqref{eq:process-phi}).
  }\label{fig:animals}
\end{figure}

\begin{figure}[h]
  \centering
  \begin{subfigure}[b]{0.3\textwidth}
    \includegraphics[width=\textwidth]{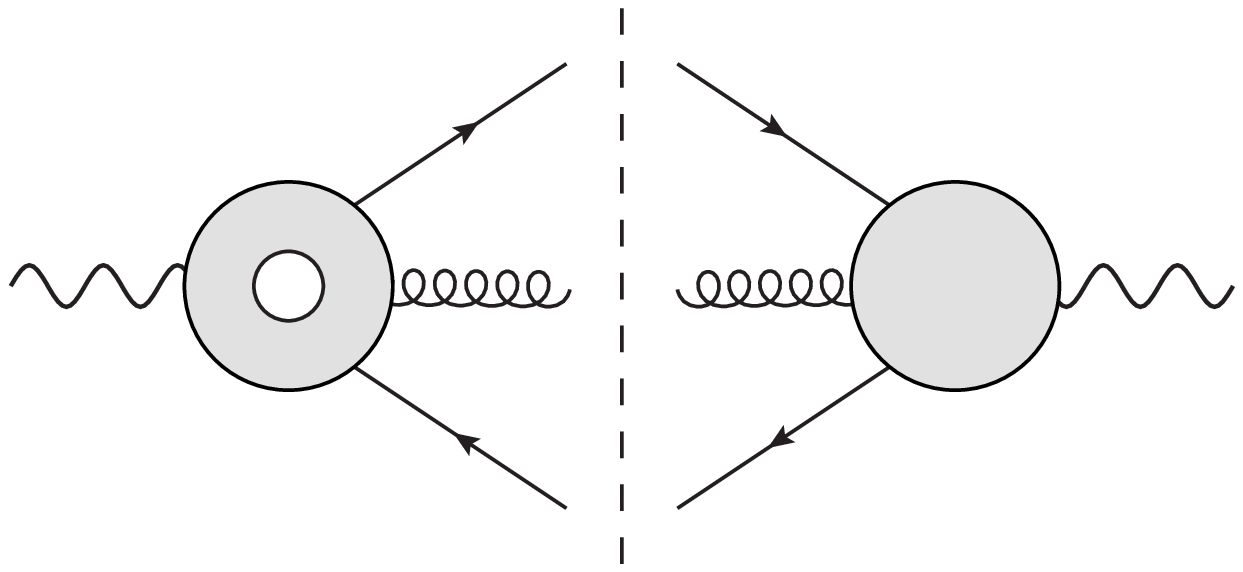}
    \caption{}
    \label{fig:amp2a}
  \end{subfigure}%
  ~
  \begin{subfigure}[b]{0.3\textwidth}
    \includegraphics[width=\textwidth]{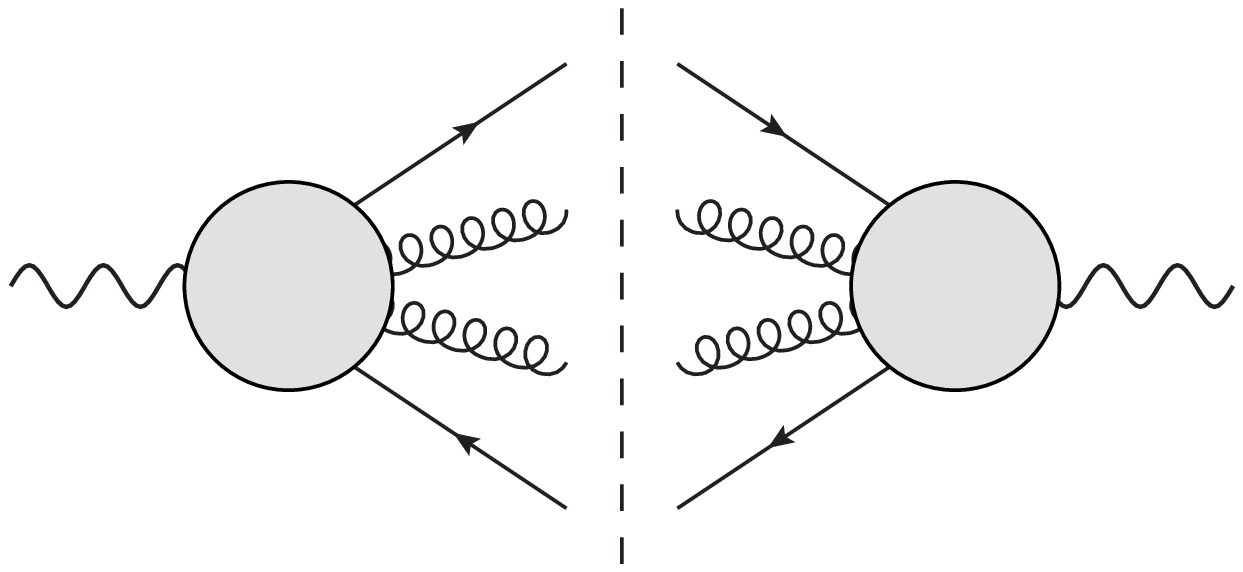}
    \caption{}
    \label{fig:amp2b}
  \end{subfigure}
  \caption{Same as fig.~\eqref{fig:animals} at NLO.}\label{fig:amp2}
\end{figure}

\begin{figure}[h]
  \centering
  \begin{subfigure}[b]{0.3\textwidth}
    \includegraphics[width=\textwidth]{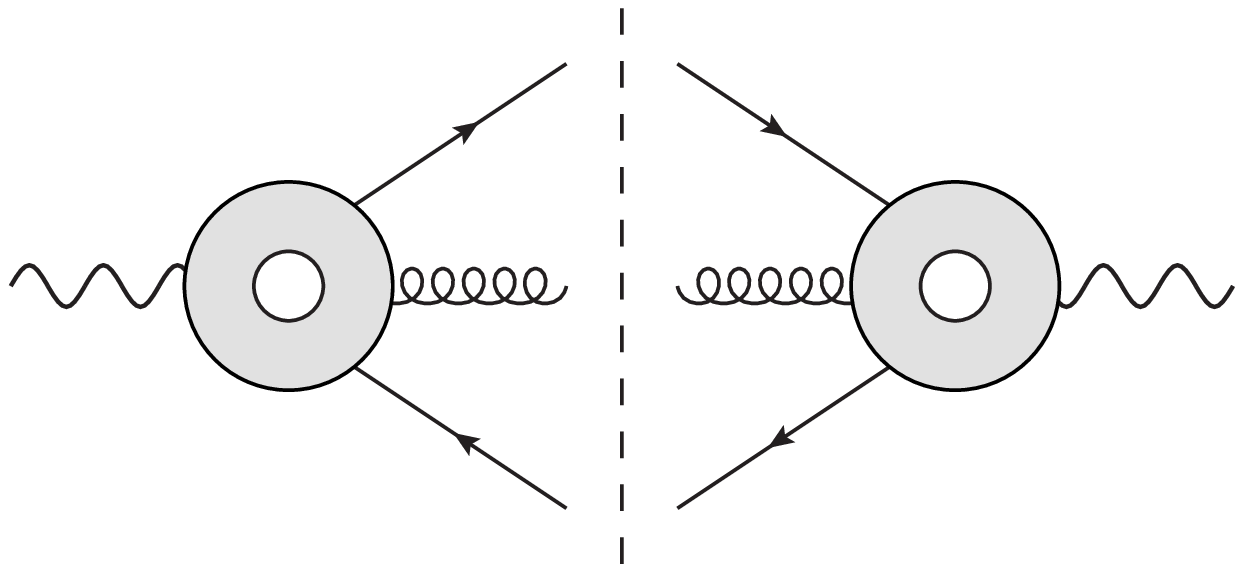}
    \caption{}
    \label{fig:amp3a}
  \end{subfigure}%
  ~
  \begin{subfigure}[b]{0.3\textwidth}
    \includegraphics[width=\textwidth]{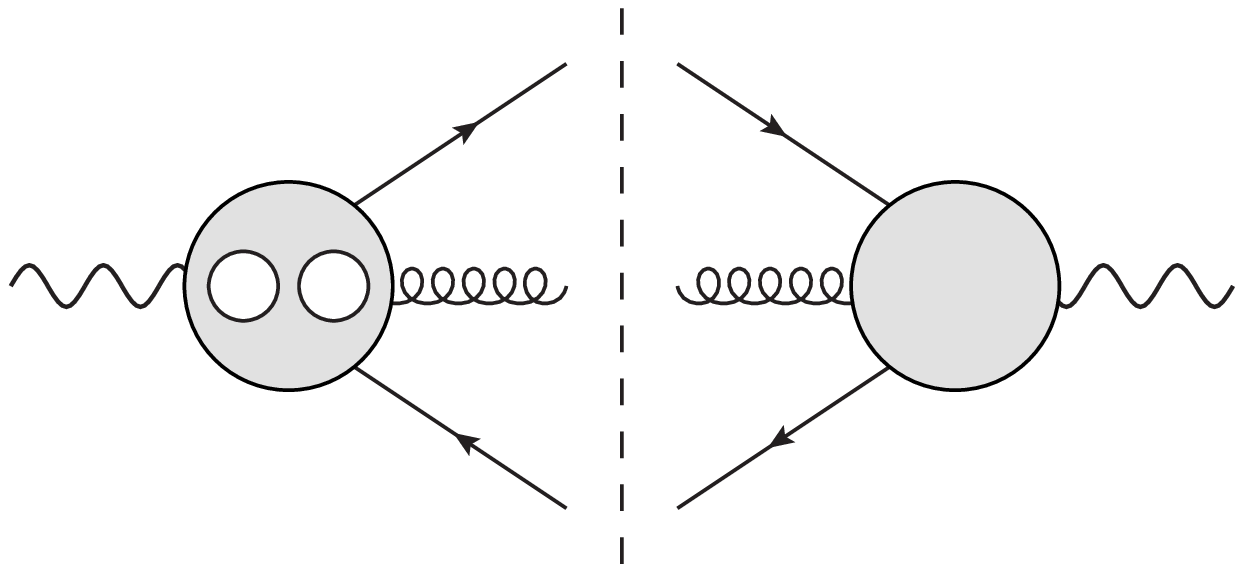}
    \caption{}
    \label{fig:amp3b}
  \end{subfigure}
  ~
  \begin{subfigure}[b]{0.3\textwidth}
    \includegraphics[width=\textwidth]{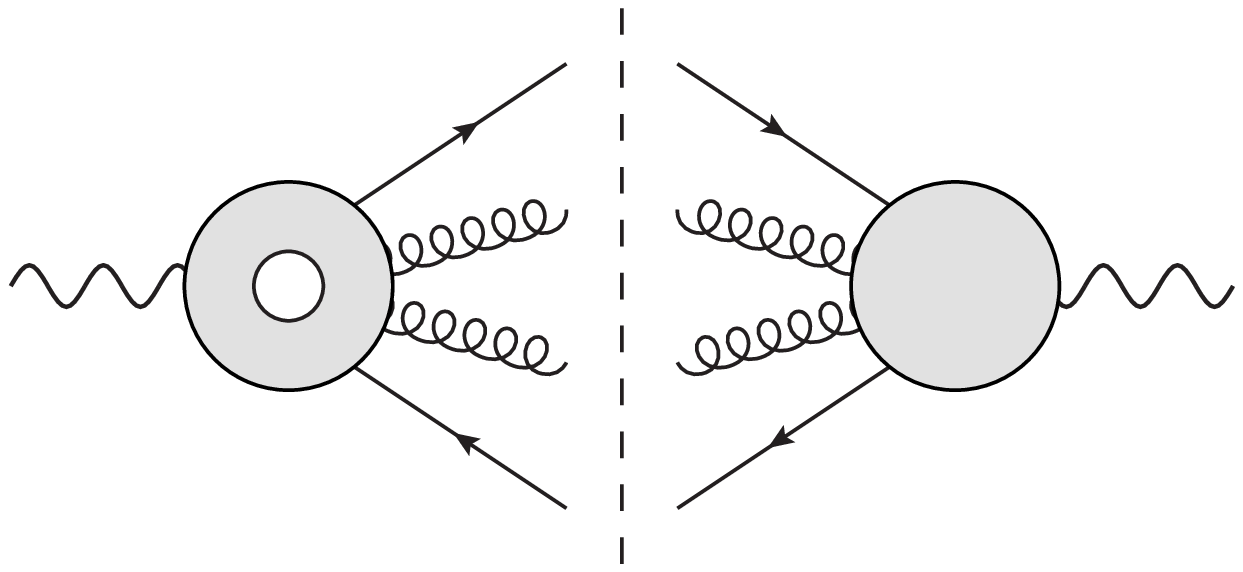}
    \caption{}
    \label{fig:amp3c}
  \end{subfigure}
  \vspace{4mm}

  \begin{subfigure}[b]{0.3\textwidth}
    \includegraphics[width=\textwidth]{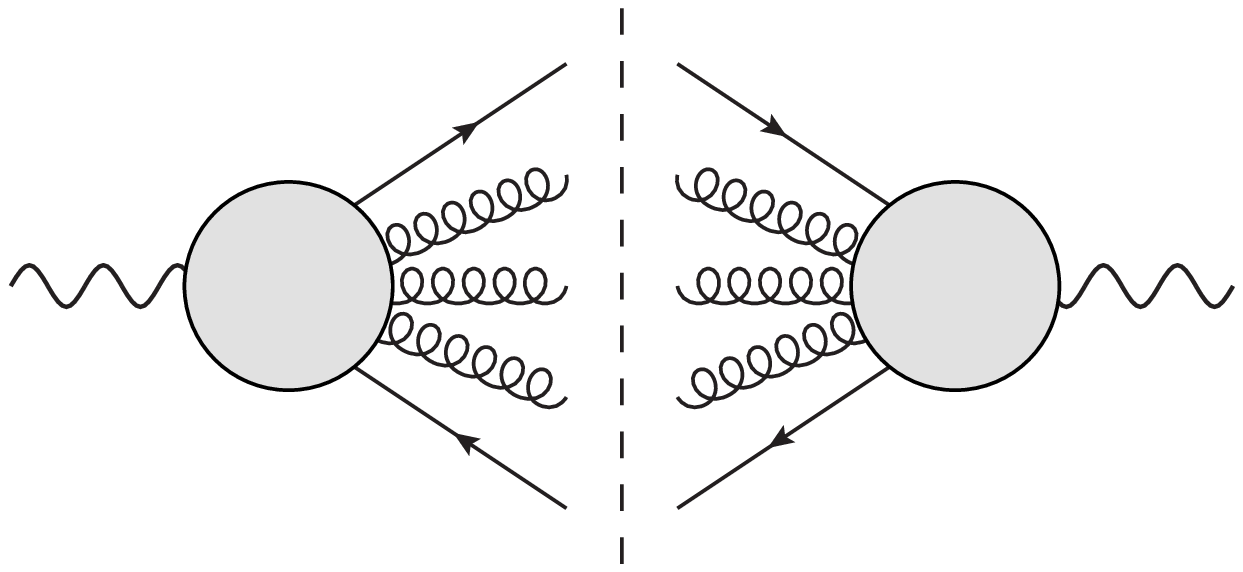}
    \caption{}
    \label{fig:amp3d}
  \end{subfigure}
  ~
  \begin{subfigure}[b]{0.3\textwidth}
    \includegraphics[width=\textwidth]{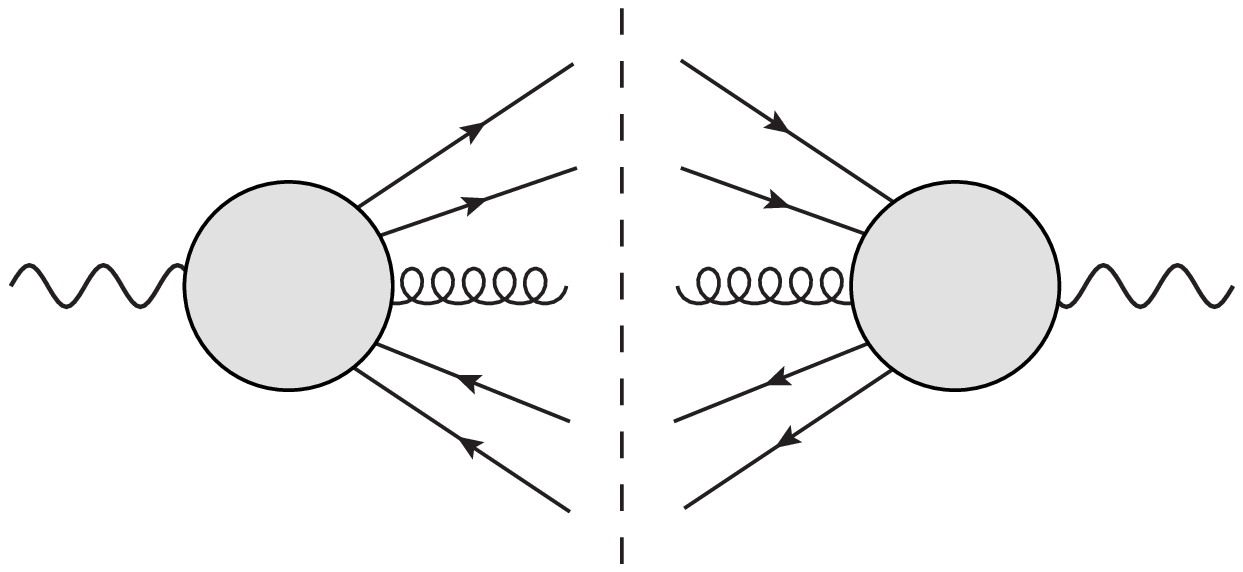}
    \caption{}
    \label{fig:amp3e}
  \end{subfigure}
  \caption{Same as fig.~\eqref{fig:animals} at NNLO.}\label{fig:amp3}
\end{figure}

As described in \cite[eq.~(38)]{Almasy:2011eq} the uncertainty for $P_{qg}^{(2)T}$ is proportional to $\beta_0$, 
which means that only contributions with closed or cut fermion loops should be considered.
For that reason, we do not need to account for the topologies depicted in fig.~\eqref{fig:amp3d}, 
which are sub-leading in $n_f$.
The two-loop contributions in fig.~(\ref{fig:amp3a}, \ref{fig:amp3b}) 
have been recently calculated in \cite{Duhr:2014nda}, while the 
the contributions in fig.~\eqref{fig:amp3c} can be constructed from one-loop
helicity amplitudes calculated in \cite{Bern:1997sc} ($\gamma^* \to 4$~partons) 
and \cite{Badger:2009hw,Badger:2009vh} ($\phi \to 4$~partons), respectively. 
For those contributions the final-state integration 
is of NLO complexity and, hence, is considered to be simple.

Finally, the contributions of fig.~\eqref{fig:amp3e} are not known 
and should be computed by explicitly performing the final-state integration.
This is a non-trivial task and in the next section we describe how to complete it.

\section{Final-state Integration} \label{sec:3}

The phase-space for the $n$-particle final-state integration in eq.~\eqref{eq:W} reads as
\begin{equation} \label{eq:psn}
    \int \D \mathrm{PS}(n)
    =
    \int \prod_{i=0}^{n} \D^m\! k_i \; \delta^+(k_i^2) \; \delta\left(x-\frac{2\sd{q}{k_0}}{q^2}\right) \; \delta(q-\sum_{j=0}^{n} k_j)
.
\end{equation}
Integrals of this type can be calculated analytically with up to $n=4$
particles in the final state using explicit parametrizations 
of the phase-space (for example, see \cite{Rijken:1996ns,GGDH03}).
However, with $n=5$ particles in the final state such an approach does not look promising.

Alternatively, we propose to find such integrals with the method of
differential equations (see, e.g., \cite{Smirnov:2004ym}), 
which has been used in a number of recent state-of-the-art computations.
The idea is as follows
\begin{enumerate}
  \item Find the integration-by-parts (\IBP) rules and reduce all the integrals to the small set of masters.
  \item Build a system of differential equations in $x$ for the masters.
  \item Choose an appropriate basis of new masters, so that the r.h.s of the
    system of equations vanishes in the limit $\eps \to 0$.
  \item Solve a new system of equations as a series in the parameter $\eps$.
  \item Find the integration constants from inclusive integrals.
\end{enumerate}
Let us briefly discuss each of the above steps and provide some examples from calculations of the contributions in fig.~\eqref{fig:amp2b}.

\subsection{Integration-by-part rules and master integrals}

\begin{figure}[h]
  \centering
  \begin{subfigure}[b]{0.175\textwidth}
    \includegraphics[width=\textwidth]{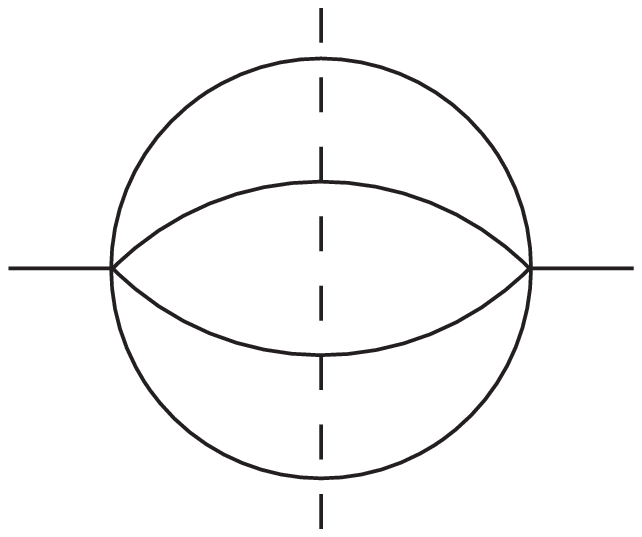}
    \caption*{$J_1 = \{\}$}
    \label{fig:mouse}
  \end{subfigure}%
  ~
  \begin{subfigure}[b]{0.175\textwidth}
    \includegraphics[width=\textwidth]{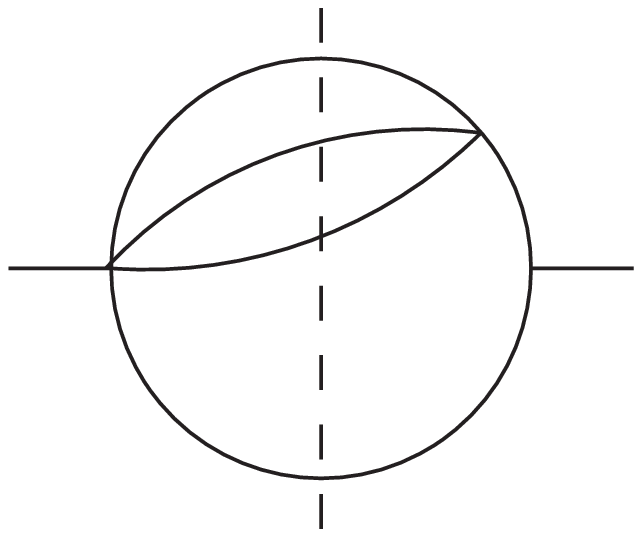}
    \caption*{$J_2 = \{1\}$}
    \label{fig:gull}
  \end{subfigure}
  ~
  \begin{subfigure}[b]{0.175\textwidth}
    \includegraphics[width=\textwidth]{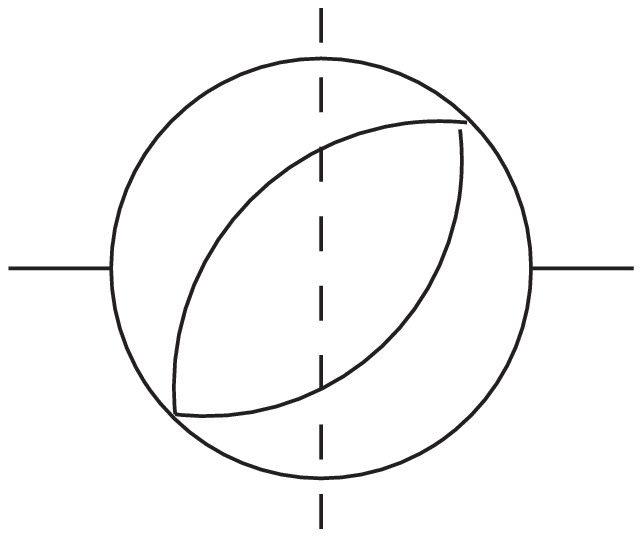}
    \caption*{$J_3 = \{1,2\}$}
    \label{fig:gull}
  \end{subfigure}
  ~
  \begin{subfigure}[b]{0.175\textwidth}
    \includegraphics[width=\textwidth]{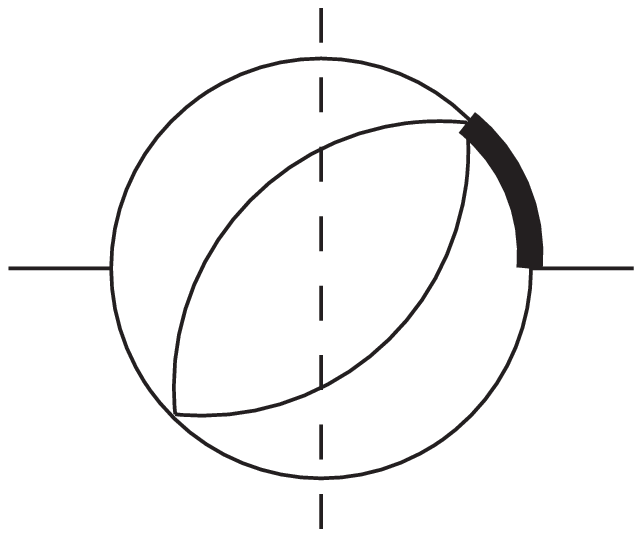}
    \caption*{$J_4 = \{1,1,2\}$}
    \label{fig:gull}
  \end{subfigure}
  ~
  \begin{subfigure}[b]{0.175\textwidth}
    \includegraphics[width=\textwidth]{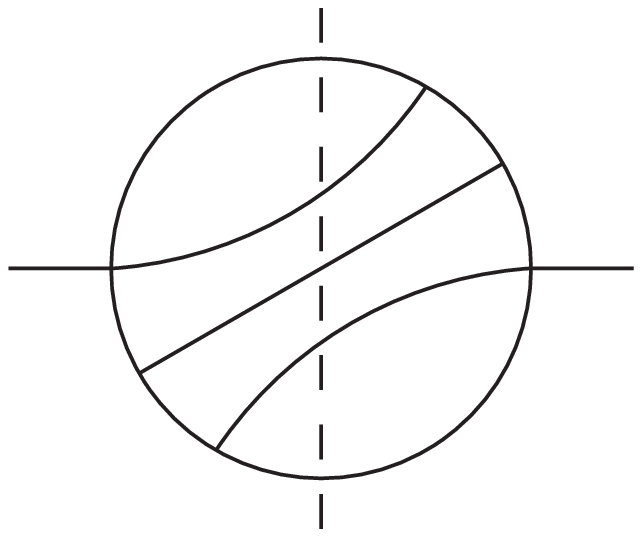}
    \caption*{$J_5 = \{1,2,3,5\}$}
    \label{fig:gull}
  \end{subfigure}
  \vspace{4mm}

  \begin{subfigure}[b]{0.175\textwidth}
    \includegraphics[width=\textwidth]{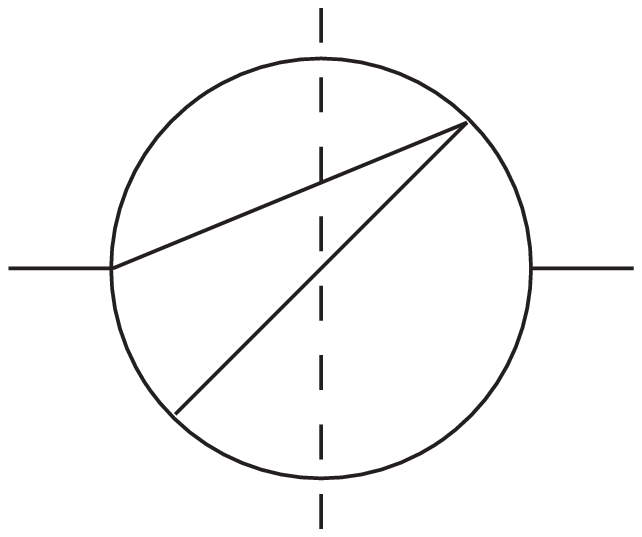}
    \caption*{$J_6 = \{2,6\}$}
    \label{fig:gull}
  \end{subfigure}
  ~
  \begin{subfigure}[b]{0.175\textwidth}
    \includegraphics[width=\textwidth]{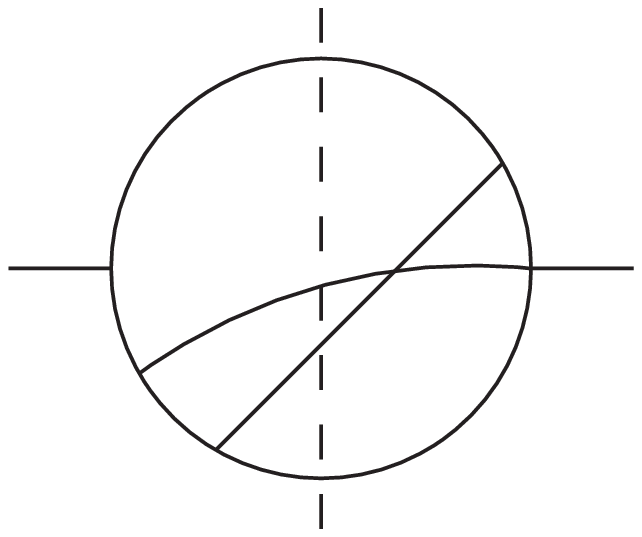}
    \caption*{$J_7 = \{2,4,6\}$}
    \label{fig:gull}
  \end{subfigure}
  ~
  \begin{subfigure}[b]{0.175\textwidth}
    \includegraphics[width=\textwidth]{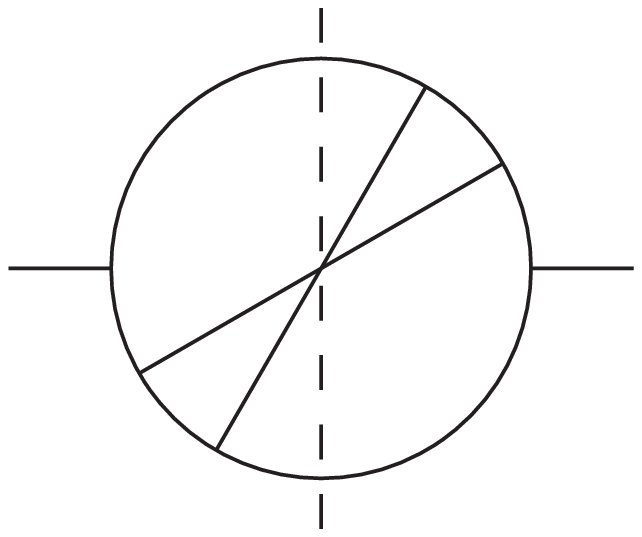}
    \caption*{$J_8 = \{1,2,6,7\}$}
    \label{fig:gull}
  \end{subfigure}
  ~
  \begin{subfigure}[b]{0.175\textwidth}
    \includegraphics[width=\textwidth]{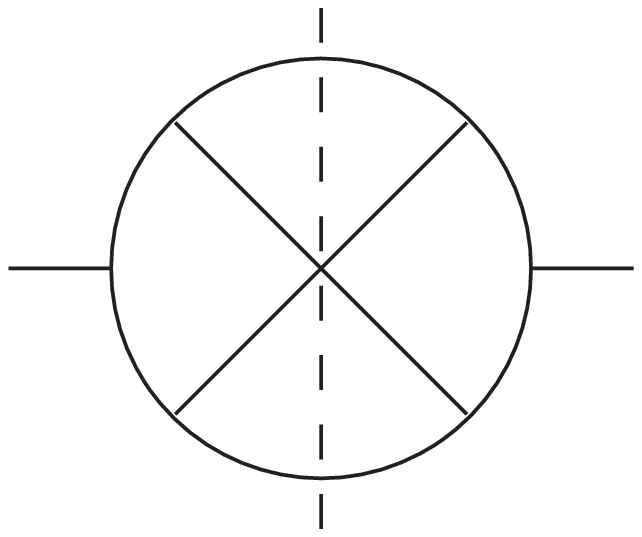}
    \caption*{$J_9 = \{3,5,6,7\}$}
    \label{fig:gull}
  \end{subfigure}
  \vspace{4mm}
  \caption{Master integrals for the real emission contributions to the
    time-like splitting functions at NLO (cf. fig.~\ref{fig:amp2b}).
    The fat line denotes a propagator raised to second power.}\label{fig:4}
\end{figure}

Integration-by-parts (IBP) \cite{CT81,Tka81} is a powerful tool to reduce a set of integrals of common structure to the small number of master integrals.
Nowadays, there are several tools to automatically generate IBP reduction
rules, e.g. \FIRE~\cite{Smi08}, \LiteRed~\cite{Lee12,Lee13} 
or \Reduze~\cite{MS12}.
We chose \LiteRed out of that list because this tool supports cut propagators, 
a feature that leads to additional simplifications in final-state integrals, 
which is essential at higher orders.
At NLO (see fig.~\eqref{fig:amp2b}) 
\LiteRed demonstrated good performance and generated all the IBP rules 
(in several hours on a standard PC) and found 9 master integrals, depicted in fig.~\ref{fig:4}.
In order to parametrize those integrals we introduce the following notation
\begin{equation}
  J_i(x,\eps)
  =
  \{a_1,\ldots,a_n\}
  =
  \int \D \mathrm{PS}(n)
    \frac{1}{P_{a_1} \ldots P_{a_n}}
,
\end{equation}
with propagators for the contributions from fig.~\eqref{fig:amp2b} defined as
\begin{equation}
\begin{aligned}
  P_1 & = (q-k_1)^2
  &
  P_2 & = (q-k_2)^2
  &
  P_3 & = (q-k_1-k_3)^2 
  &
  P_4 & = (q-k_1-k_2)^2 
  \\
  P_5 & = (q-k_2-k_3)^2
  &
  P_6 & = (k_2+k_3)^2 
  &
  P_7 & = (k_1+k_3)^2.
\end{aligned}
\end{equation}

It is worth to mention that this set of masters differs from the one in Mellin space calculated for the same contribution, see~\cite{MM06}.

For the contributions from fig.~\eqref{fig:amp3e} \LiteRed can solve the major part of IBP sectors. 
However for a few it takes months of CPU time with no guarantee of completion.
Preliminary attempts show that these sectors can be solved with a private
version of \Reduze, which also features support for cut propagators.

\subsection{Differential equations and solutions for master integrals}

The master integrals from fig.~\eqref{fig:4} may be integrated explicitly, which at higher orders becomes a challenging task.
A simpler approach is to solve a linear system of differential equations, which is constructed as
\begin{equation} \label{eq:mde}
  \frac{\partial J_i(x,\eps)}{\partial x} = \sum_{j=1}^n A_{ij}(x,\eps) J_j(x,\eps), \quad i,j=1\ldots n\, ,
\end{equation}
where matrix $A_{ij}(x,\eps)$ is obtained by reducing the l.h.s. of eq.~\eqref{eq:mde} using IBP reduction rules.
In the case of masters from fig.~\eqref{fig:4} we obtain the following matrix
{\scriptsize
\begin{equation*}
\hspace{-9mm}
\left(\hspace{-2mm}
\begin{array}{ccccccccc}
 \frac{(2 \eps-1) (2 x-1)}{ (1-x) x} \hspace{-2mm}&\hspace{-2mm} 0 \hspace{-2mm}&\hspace{-2mm} 0 \hspace{-2mm}&\hspace{-2mm} 0 \hspace{-2mm}&\hspace{-2mm} 0 \hspace{-2mm}&\hspace{-2mm} 0 \hspace{-2mm}&\hspace{-2mm} 0 \hspace{-2mm}&\hspace{-2mm} 0 \hspace{-2mm}&\hspace{-2mm} 0 \\
 \frac{3 \eps-2}{(1-x) x} \hspace{-2mm}&\hspace{-2mm} -\frac{3 \eps-1}{ x} \hspace{-2mm}&\hspace{-2mm} 0 \hspace{-2mm}&\hspace{-2mm} 0 \hspace{-2mm}&\hspace{-2mm} 0 \hspace{-2mm}&\hspace{-2mm} 0 \hspace{-2mm}&\hspace{-2mm} 0 \hspace{-2mm}&\hspace{-2mm} 0 \hspace{-2mm}&\hspace{-2mm} 0 \\
 0 \hspace{-2mm}&\hspace{-2mm} \frac{(2 \eps-1) (3 \eps-1)}{{ \eps} (x-1) x} \hspace{-2mm}&\hspace{-2mm} \frac{2 \eps}{ 1-x} \hspace{-2mm}&\hspace{-2mm} 0 \hspace{-2mm}&\hspace{-2mm} 0 \hspace{-2mm}&\hspace{-2mm} 0 \hspace{-2mm}&\hspace{-2mm} 0 \hspace{-2mm}&\hspace{-2mm} 0 \hspace{-2mm}&\hspace{-2mm} 0 \\
 \frac{(2 \eps-1) (3 \eps-2) (3 \eps-1)}{2 { \eps} (x-1) x^2} \hspace{-2mm}&\hspace{-2mm} \frac{(2 \eps-1) (3 \eps-1) \left(x^2-10 x+1\right)}{2 (1-x) x^2 (x+1)} \hspace{-2mm}&\hspace{-2mm} 0 \hspace{-2mm}&\hspace{-2mm} \frac{2 \eps \left(x^2-3 x-2\right)}{ (1-x) x (x+1)} \hspace{-2mm}&\hspace{-2mm}  \frac{2 \eps (6 \eps-1)}{(1-x) x} \hspace{-2mm}&\hspace{-2mm} 0 \hspace{-2mm}&\hspace{-2mm} 0 \hspace{-2mm}&\hspace{-2mm} 0 \hspace{-2mm}&\hspace{-2mm} 0 \\
 0 \hspace{-2mm}&\hspace{-2mm} \frac{(2 \eps-1) (3 \eps-1)}{{ \eps} (x-1) x} \hspace{-2mm}&\hspace{-2mm} 0 \hspace{-2mm}&\hspace{-2mm} \frac{2}{x-1} \hspace{-2mm}&\hspace{-2mm} \frac{6 \eps-1}{ 1-x} \hspace{-2mm}&\hspace{-2mm} 0 \hspace{-2mm}&\hspace{-2mm} 0 \hspace{-2mm}&\hspace{-2mm} 0 \hspace{-2mm}&\hspace{-2mm} 0 \\
 \frac{4 (2 \eps-1) (3 \eps-2) (3 \eps-1)}{{ \eps^2} (1-x) x^3 (x+1)} \hspace{-2mm}&\hspace{-2mm} \frac{4 (2 \eps-1) (3 \eps-1) \left(x^2-x+1\right)}{{ \eps} (x-1) x^3 (x+1)^2} \hspace{-2mm}&\hspace{-2mm} 0 \hspace{-2mm}&\hspace{-2mm} \frac{4 \left(x^2+1\right)}{(x-1) x^2 (x+1)^2} \hspace{-2mm}&\hspace{-2mm} \frac{2 (6 \eps-1)}{(1-x) x^2 (x+1)} \hspace{-2mm}&\hspace{-2mm} \frac{(2 \eps+1) (2 x+1)}{  -x (x+1)} \hspace{-2mm}&\hspace{-2mm} 0 \hspace{-2mm}&\hspace{-2mm} 0 \hspace{-2mm}&\hspace{-2mm} 0 \\
 \frac{2 (2 \eps-1) (3 \eps-2) (3 \eps-1)}{{ \eps^2} (x-1)^2 x^2} \hspace{-2mm}&\hspace{-2mm} -\frac{(2 \eps-1) (3 \eps-1)}{{ \eps} (x-1) x^2} \hspace{-2mm}&\hspace{-2mm} \frac{2 \eps}{(x-1) x} \hspace{-2mm}&\hspace{-2mm} 0 \hspace{-2mm}&\hspace{-2mm} 0 \hspace{-2mm}&\hspace{-2mm} 0 \hspace{-2mm}&\hspace{-2mm} \frac{4 \eps+1}{ -x} \hspace{-2mm}&\hspace{-2mm} 0 \hspace{-2mm}&\hspace{-2mm} 0 \\
 \frac{2 (1-2 \eps) (3 \eps-2) (3 \eps-1)}{{ \eps^2} (x-1)^2 x^3} \hspace{-2mm}&\hspace{-2mm} \frac{2 (2 \eps-1) (3 \eps-1) (3 x-1)}{{ \eps} (x-1)^3 x^3} \hspace{-2mm}&\hspace{-2mm} \frac{4 \eps}{(1-x)^2 x} \hspace{-2mm}&\hspace{-2mm} \frac{4 \left(x^2+1\right)}{(x-1)^3 x^2} \hspace{-2mm}&\hspace{-2mm} \frac{2 (6 \eps-1) (x+1)}{(1-x)^3 x^2} \hspace{-2mm}&\hspace{-2mm} 0 \hspace{-2mm}&\hspace{-2mm} 0 \hspace{-2mm}&\hspace{-2mm} \frac{(2 \eps+1) (2 x-1)}{ (1-x) x} \hspace{-2mm}&\hspace{-2mm} 0 \\
 0 \hspace{-2mm}&\hspace{-2mm} 0 \hspace{-2mm}&\hspace{-2mm} 0 \hspace{-2mm}&\hspace{-2mm} 0 \hspace{-2mm}&\hspace{-2mm} 0 \hspace{-2mm}&\hspace{-2mm} 0 \hspace{-2mm}&\hspace{-2mm} 0 \hspace{-2mm}&\hspace{-2mm} 0 \hspace{-2mm}&\hspace{-2mm} \frac{(2 \eps+1) (2 x-1)}{ (1-x) x} \\
\end{array}
\hspace{-2mm}\right)
\end{equation*}
}%
Solutions of this system can be found in terms of harmonic polylogarithms (HPL), 
implemented in the \harmpol \cite{RV99} package for \FORM or in the \HPL \cite{Mai05} package for \Mathematica.

\subsection{Boundary conditions for master integrals}

The last step in solving for masters is to fix integration constants which appear in solutions of differential equations.
Such constants can be determined conveniently from the inclusive integrals calculated in \cite{GGDH03,MM06}.

\begin{figure}[h]
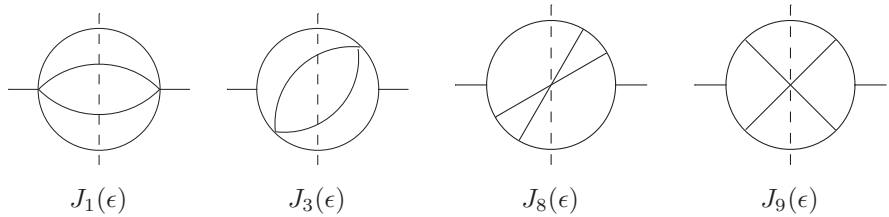

  \centering
  \begin{subfigure}[b]{0.175\textwidth}
    \includegraphics[width=\textwidth]{img/j1.eps}
    \caption*{$J_1(\eps)$}
    \label{fig:mouse}
  \end{subfigure}%
  ~
  \begin{subfigure}[b]{0.175\textwidth}
    \includegraphics[width=\textwidth]{img/j2.eps}
    \caption*{$J_3(\eps)$}
    \label{fig:gull}
  \end{subfigure}
  ~
  \begin{subfigure}[b]{0.185\textwidth}
    \includegraphics[width=\textwidth]{img/j5.eps}
    \caption*{$J_8(\eps)$}
    \label{fig:gull}
  \end{subfigure}
  ~
  \begin{subfigure}[b]{0.185\textwidth}
    \includegraphics[width=\textwidth]{img/j6.eps}
    \caption*{$J_9(\eps)$}
    \label{fig:gull}
  \end{subfigure}
  \vspace{4mm}
  \vspace{4mm}
  \caption{Masters for the phase-space integrals at NLO}\label{fig:5}
\end{figure}

In the inclusive case, there are only 4 master integrals (fig.~\ref{fig:5}),
hence IBP rules exist in order to reduce the 9 masters integrated in the $x$ variable.
\begin{equation}
  J_i(\eps) = \int_0^1 \D x \; J_i(x,\eps)\, ,
\end{equation}
which is basically the first Mellin moment.

Such rules can be found with the help of \LiteRed and read
\begin{equation}
\begin{aligned}
  J_2(\eps) & = \frac{3-4\eps}{1-2\eps} J_1(\eps)\, ,
  \qquad
  J_4(\eps) = -\frac{(2-3\eps)(3-4\eps)}{\eps} J_1(\eps) - (1-3\eps) J_3(\eps)\, ,
  \qquad
  J_5(\eps) = J_8(\eps)\, ,
  \\
  J_6(\eps) & = -\frac{2(2-3\eps)(3-4\eps)}{\eps(1-2\eps)} J_2(\eps)\, ,
  \qquad
  J_7(\eps) = \frac{(1-3 \eps) (1-4 \eps) (2-3 \eps) (3-4 \eps)}{2 \eps^4} J_1(\eps)\, ,
\end{aligned}
\end{equation}
which are exact in $m$ dimensions, i.e., to all orders in $\eps$.

\section{Summary}  \label{sec:4}

In these proceedings we have motivated the necessity 
for a NNLO QCD computation of the time-like splitting functions from first principles.
We have outlined a strategy how to calculate the required three-loop contributions. 
Based on the physical process $\process$, we have analyzed the transverse fragmentation function 
$\Ft(x,\eps)$ in $x$-space using dimensional regularization.
The most challenging part of our approach lies in the evaluation of the final-state integrals with a projection to $x$-space. 
For the NNLO case, this corresponds to four-loop integrals with one massive leg and cut internal propagators.
Thanks to an analysis in Mellin space and to partial results available in the literature, 
the number of topologies, and hence the integrals to calculate, can be largely reduced.
Despite such simplifications there are, however, still about 80 integrals left to calculate and 
we employ the integration-by-parts technique and the method of differential equations to obtain the master integrals.
The latter method has become widely used in the last years and we have
described its application to our calculations.

\subsection*{Acknowledgments}
This work has been supported by the Research Executive Agency (REA) with the European Union Grant PITN-GA-2010-264564 (LHCPhenoNet).
The work of O.G. has been partly supported by Narodowe Centrum Nauki with the Sonata Bis grant DEC-2013/10/E/ST2/00656.
O.G. thanks for the hospitality of the Theory Group of the University of Hamburg where major part of this research was done.

\newpage

\providecommand{\href}[2]{#2}\begingroup\raggedright\endgroup

\end{document}